\documentclass[fleqn,twoside]{article}
\usepackage{espcrc2}

\usepackage{graphicx}
\usepackage[figuresright]{rotating}
\usepackage{amsmath}

\def\gsim{\raise0.3ex\hbox{$>$\kern-0.75em\raise-1.1ex\hbox{$\sim$}}}
\def\lsim{\raise0.3ex\hbox{$<$\kern-0.75em\raise-1.1ex\hbox{$\sim$}}}

\title{WMAP, Planck, cosmic rays and unconventional cosmologies}

\author{Luis Gonzalez-Mestres\address{LAPP, Universit\'e de Savoie, CNRS/IN2P3, B.P. 110, 74941 Annecy-le-Vieux Cedex, France}}

\begin{document}

\begin{abstract}
The claim by Gurzadyan et al. that the cosmological sky is a weakly random one where "the random perturbation is a minor component of mostly regular signal" has given rise to a series of useful exchanges. The possibility that the Cosmic Microwave Background radiation (CMB) data present trends in this direction would have strong implications for unconventional cosmologies. Similarly, data on ultra-high energy cosmic rays may contain signatures from new Physics generated beyond the Planck scale. It therefore seems legitimate, from a phenomenological point of view, to consider pre-Big Bang cosmologies as well as patterns where standard particles would not be the ultimate constituents of matter and the presently admitted principles of Physics would not necessarily be the fundamental ones. We discuss here prospects for some noncyclic, nonstandard cosmologies. 
\vspace{1pc}
\end{abstract}

\maketitle

\section{Introduction}

Is the Planck scale the ultimate fundamental scale, or is there new Physics beyond the Planck scale ? Is the standard Big Bang scenario the ultimate cosmological theory, or should it be modified taking into account a not yet explored pre-Big Bang era ? Are string patterns \cite{Mukhi} the ultimate description of matter, or are they actually reminiscent of an underlying composite structure ? Can we experimentally find signatures from such a possible new Physics and Cosmology ? 

The work by Gurzadyan et al. on CMB randomness \cite{Gurzadyan1}, together with the paper by Gurzadyan and Penrose \cite{GurzadyanPenrose1} considering the possibility that concentric circles in our CMB sky provide a signature of pre-Big bang black-hole encounters in a conformal cyclic cosmology (CCC), has led to an interesting and useful controversial debate \cite{Gurzadyan2,GurzadyanPenrose2}. But whatever the final conclusion, it would seem quite natural that possible pre-Big Bang physics manifests itself through WMAP \cite{WMAP} and Planck \cite{Planck} data.

These considerations apply also to possible noncyclic cosmologies involving real new Physics beyond the Planck scale \cite{Gonzalez-Mestres1,Gonzalez-Mestres2}, that can potentially lead to signatures in WMAP and Planck data. Using these signatures, it would be possible to get some insight on Physics between the Planck scale and an ultimate fundamental scale. 

Similarly, the properties of ultra-high energy (UHE) cosmic rays \cite{Auger,Hires} may contain effects generated at an energy scale beyond Planck energy. A simple example can be obtained taking for a ultra-high energy particle the quadratically deformed dispersion relation \cite{Gonzalez-Mestres1,Gonzalez-Mestres3}: 
\begin{equation}
\begin{split}
E ~ \simeq ~ p~c~+~m^2~c^3~(2~p)^{-1}~ \\
-~p~c~\alpha ~(p~ c ~E_a^{-1})^2/2~~~~~
\end{split}
\end{equation}
\noindent
where $E$ is the energy, $p$ the momentum, $c$ the speed of light, $m$ the mass, $\alpha $ a constant standing for the deformation strength and $E_a$ the fundamental energy scale at which the deformation is generated. Then, the deformation term $\Delta ~E~\simeq ~-~p~c~\alpha ~(p~ ~c ~E_a^{-1})^2/2$ equals the mass term $m^2~c^3~(2~p)^{-1}$ at a transition energy $E_{trans}$ given by :
\begin{equation}
E_{trans} ~ \simeq ~ \alpha ^{-1/4}~(E_a~m)^{1/2}~c
\end{equation}
For a proton, and taking $\alpha $ = 1, one would have $E_{trans} $ = 5 x $10^{19}$ eV for $E_a ~\simeq ~$ 2.8 x $10^{21}$ GeV, more than 200 times the Planck energy $E_{Planck}$. A similar equation for a $10^{20}$ eV proton would yield $E_a ~\simeq ~ 10^3 ~ E_{Planck}$.

Therefore, there exists serious motivation to explore possible pre-Big Bang and pre-Planck physics and cosmologies. In what follows, we discuss some proposals in this direction.
 
\section{A new space-time geometry ?}

Standard quantum field theory has been quite successfully formulated using four real space-time dimensions and the standard Lorentz group.
 
However, such a representation of space-time cannot really incorporate spin-1/2 particles, which belong to a representation of the covering group SL(2,C). Although this has not been a major problem to formulate quantum field theory, full consistency of the picture would require that spin-1/2 particles be a direct representation of the space-time symmetry used. The problem remains if only space rotations are considered and a preferred reference frames is used, in which case the relevant covering group is SU(2).

For this reason, we suggested in 1996-97 \cite{Gonzalez-Mestres4} that space-time be described by two complex spinorial coordinates instead of the usual Lorentz real quadrivector. Then, given a spinor $\xi $, and considering the positive SU(2) scalar $\mid \xi \mid ^2$ $=$ $\xi ^\dagger \xi $ where the dagger stands for hermitic conjugate, it is possible to define a positive cosmic time $t~=~\mid \xi \mid$ in the (preferred ?) reference frame used. In this case, there would be a natural origin of space and time, $\xi ~=~0$. This leads to a naturally expanding Universe where the space at constant time $t_0$ is given by the $S^3$ hypersphere $\mid \xi \mid~=~t_0$. Space translations are described by cosmic SU(2) transformations acting on the constant-time spinor hypersphere, whereas space rotations are local SU(2) transformations acting on the translations (see the Post Scriptum to \cite{Gonzalez-Mestres1}).

More precisely, if $\xi _0$ is the observer position on the $\mid \xi \mid $ = $t_0$ hypersphere, one can write for a point $\xi $ of the same spatial hypersphere :
\begin{equation}
\xi ~=~ U ~\xi _0
\end{equation}
where $U$ is the SU(2) matrix :
\begin{equation}
U~=~exp~(i/2~~t_0^{-1}~{\vec \sigma }.{\vec {\mathbf x}})~
\equiv U ({\vec {\mathbf x}}) 
\end{equation}
and ${\vec \sigma }$ is the vector formed by the Pauli matrices. The vector ${\vec {\mathbf x}}$ can be interpreted as the spatial position vector of $\xi $ with respect to $\xi _0$ at constant time $t_0$. 

Then, a standard space rotation around $\xi _0 $ is defined by the action of a SU(2) element $U({\vec {\mathbf y}})$ turning any $U({\vec {\mathbf x}})$ into $U({\vec {\mathbf y}}) ~ U({\vec {\mathbf x}}) ~ U({\vec {\mathbf y}})^\dagger $. The vector $\vec {\mathbf y}$ provides the rotation axis and angle. 

Cosmologically comoving frames correspond to spinorial straight lines through $\xi ~=~0$, whereas straight lines through $\xi _0 $ describe conventional inertial frames.

Such a spinorial space-time automatically yields a ratio between relative velocities and distances at cosmic scale similar to standard cosmology \cite{Lemaitre,Hubble} and equal to the inverse of the age of the Universe (the only available scale at that stage). This result, similar to those by Hubble and Lema{\^i}tre, is obtained here on purely geometric grounds without introducing standard matter, relativity, gravitation or even specific space units. It is therefore tempting to conjecture that the use of a spinorial space-time leads to a simple solution of the cosmological constant problem and provides a natural alternative to conventional inflation, dark mater and dark energy scenarios.  

Thus, as the present experimental value of the expansion rate of the Universe is close to that predicted by our space-time geometry, it seems worth exploring the possibility that recent data and analysis on the acceleration of the expansion of the Universe \cite{DEObservations,Daly} actually describe a fluctuation generated by matter and gravity in the past, but not a long-term trend for the future. Gravitation would generate curved trajectories in the spinorial space-time and deform the expansion of standard matter. In this context, the need for dark energy appears far from obvious.

As the spinorial description of space-time seems particularly well adapted to account for the large-scale properties of the Universe, the question of its link to matter needs to be raised. Since the orbital motion of standard matter cannot generate half-integer spins, possible new physics beyond Planck scale must be considered.

\section{Unconventional preon models}

As stressed in \cite{Gonzalez-Mestres1,Gonzalez-Mestres5}, the string picture originated from the dual resonance models of hadronic physics \cite{DRM1}, and was initially interpreted \cite{DRM2} in terms of "fishnet" Feynman diagrams involving quark and gluon lines. Similarly, present string models may actually be the expression of an underlying composite structure of standard matter. An important question is therefore that of the possible properties of the constituents of conventional particles, including their implications for the structure of the physical vacuum. 

Although the first preon models \cite{preons} where based on a "building block" picture without modifying the space-time geometry seen by the new ultimate constituents, the superbradyon pattern proposed in 1995 \cite{Gonzalez-Mestres6} introduced a radical change, suggesting that the critical speed of such constituents be much larger than that of light, just as the critical speed of standard particles is much larger than that of phonons in a solid. The standard particles would actually be excitations of a medium (the vacuum) made of more fundamental matter. Then, Lorentz symmetry and conventional quantum field theory would be just low-energy limits. 

Superbradyons (superluminal preons) provide an illustrative example of how new physics beyond Planck scale may differ from standard particle physics. As stressed in \cite{Gonzalez-Mestres1,Gonzalez-Mestres5}, standard relativity is far from being the only fundamental principle that can be questioned near the Planck scale or beyond it. Quantum mechanics or the conventional energy and momentum conservation may also fail at such scales.

Then, symmetries will not necessarily become more and more exact as the energy scale increases and masses can be neglected. It may happen instead that this behaviour of the laws of Physics holds only below some critical transition energy scale. Above this transition energy, observations would start being sensitive to new features of the particle internal structure and to properties of the real fundamental matter beyond Planck scale. The energy scale $E_trans$ in (2) provides a simple example of such a transition in the case of special relativity and particle kinematics.

As suggested in the Post Scripta to \cite{Gonzalez-Mestres2}, if the fundamental state of matter is superbradyonic, the Higgs boson does not need to be permanently materialized in vacuum, and similarly for the zero modes of the bosonic harmonic oscillators. It may happen that the relevant processes of standard quantum field theory be dynamically generated at the relevant frequencies by the superbradyonic vacuum only in the presence of conventional matter. This would in principle solve the cosmological constant problem. One may also expect modifications of the high-energy internal lines of Feynman diagrams when the energies of the virtual particles are much higher than those involved in the process considered. 

Another issue \cite{Gonzalez-Mestres1} concerns standard causality and the relevance of space-like distances measured on a $S^3$ hypersphere. Between the above considered $\xi _0$ and $\xi ~=~ U ~ \xi _0$, the position vector ${\vec {\mathbf x}}$ is not the only way to relate the two points. The "direct" spinorial separation $\Delta \xi ~= ~\xi ~-~ \xi _0$ can also be considered, but the points of the spinorial straight line belong to the past as compared to $\xi _0$ and $\xi $. To explain the generation of spin-1/2 particles, one may assume that causality does no longer hold below a critical distance scale (Planck ?), and that $\Delta \xi $ describes then the actual physical position of $\xi $ with respect to $\xi _0$. Thus, superbradyonic physics would naturally generate the half-integer internal angular momenta of standard particles.

Superbradyons and similar objects may exist in our Universe as free particles, in which case they are expected to emit "Cherenkov" radiation in the form of standard particles \cite{Gonzalez-Mestres6} until their speed becomes close to $c$. They can then form a cosmological sea \cite{Gonzalez-Mestres1,Gonzalez-Mestres2} and possibly be part of the dark matter. They would also modify the history of the early universe, provide an alternative to inflation and possibly leave signatures observable in WMAP and Planck data. As the superbradyon rest energy is $m ~ c_s^2$, where $c_s ~\gg ~c$ is the superbradyon critical speed, one may expect spontaneous superbradyon decays to produce very high energy cosmic rays.  

\section{Conclusion and prospects}

There exist serious reasons to pay attention to the possible detectable effects of noncyclic pre-Big Bang cosmologies and of the new Physics beyond the Planck scale that may be associated to these cosmologies. Such a task does not appear impossible. WMAP and Planck data, but also UHE cosmic-rays experiments, may provide relevant signatures.

The spinorial space-time geometry discussed here appears particularly well suited to provide the framework for a new theory of matter and of its ultimate constituents. Further work is needed in this direction.

\end{document}